\documentclass[aps,twocolumn,showpacs,floatfix,superscriptaddress]{revtex4-1}
\usepackage[linktocpage,bookmarksopen,bookmarksnumbered]{hyperref}
\usepackage[utf8]{inputenc} 
\usepackage{dcolumn}
\usepackage{ulem}
\usepackage{graphicx}
\usepackage{chemformula}
\usepackage{amsfonts}
\usepackage{amsmath,bm,amssymb}
\usepackage{blkarray}
\usepackage{braket} 
\usepackage{gensymb}

\begin{document}

\title{Reliability and applicability of magnetic force linear response theory:
	Numerical parameters, predictability, and orbital resolution}

\author{Hongkee Yoon}
\affiliation{Department of Physics, KAIST, 291 Daehak-ro, Yuseong-gu, Daejeon 34141, Republic of Korea }

\author{Taek Jung Kim} 
\affiliation{Department of Physics, KAIST, 291 Daehak-ro, Yuseong-gu, Daejeon 34141, Republic of Korea }
\author{Jae-Hoon Sim}
\author{Seung Woo Jang}
\affiliation{Department of Physics, KAIST, 291 Daehak-ro, Yuseong-gu, Daejeon 34141, Republic of Korea }

\author{Taisuke Ozaki} 
\affiliation{Institute for Solid State Physics, The University of Tokyo,
	Kashiwa 277-8581, Japan }

\author{Myung Joon Han} \email{mj.han@kaist.ac.kr}
\affiliation{Department of Physics, KAIST, 291 Daehak-ro, Yuseong-gu, Daejeon 34141, Republic of Korea }
\affiliation{ KAIST Institute for the NanoCentury, Korea Advanced
  Institute of Science and Technology, Daejeon 305-701, Korea }

\date{\today}

\begin{abstract}
We investigated the reliability and applicability of so-called magnetic force linear response method  to calculate spin-spin interaction strengths from first-principles. We examined the dependence on the numerical parameters including the number of basis orbitals and their cutoff radii within non-orthogonal LCPAO (linear combination of pseudo-atomic orbitals) formalism. It is shown that the parameter dependence and the ambiguity caused by these choices are small enough in comparison to the other computation approach and experiments. Further, we tried to pursue the possible extension of this technique to a wider range of applications. We showed that magnetic force theorem can provide the reasonable estimation especially for the case of strongly localized moments even when the ground state configuration is unknown or the total energy value is not accessible. The formalism is extended to carry the orbital resolution from which the matrix form of the magnetic coupling constant is calculated. From the applications to Fe-based superconductors including LaFeAsO, NaFeAs, BaFe$_2$As$_2$ and FeTe, the distinctive characteristics of orbital-resolved interactions are clearly noticed in between single-stripe pnictides and double-stripe chalcogenides.
\end{abstract}


\maketitle

\section{Introduction}

\begin{figure}[!th]
    \begin{center}
    \includegraphics[width=0.95\linewidth,angle=0]{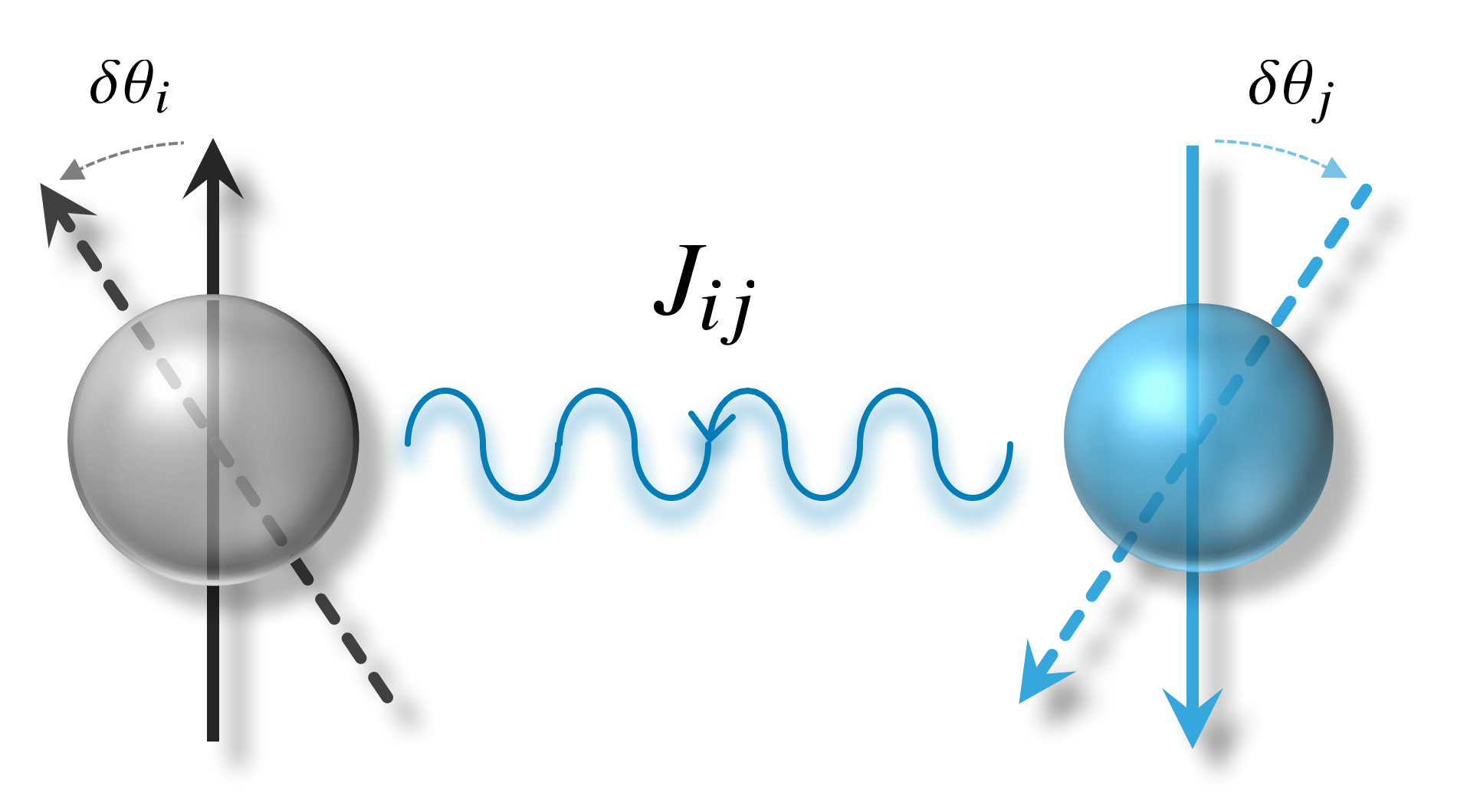}
    \caption{
(Color online) Schematic picture for illustrating MFT. MFT estimates $J$ as an energy response 
to a small spin angle tilting ($\delta\theta_i, \delta\theta_j$) in a perturbative manner.
    \label{Figure_1}}
    \end{center}
\end{figure}

Magnetic interactions in solids are often described in terms of effective spin Hamiltonian, $H=-J~{\bf S}_1\cdot{\bf S}_2$, and the determination of $J$ is the key step to study magnetism and related phenomena such as unconventional superconductivity. While several different experimental techniques can be utilized to measure the magnetic interaction parameter, a notable feature of first-principles calculation is that the independent estimation of $J$ is feasible. One typical way is to compare the calculated total energies corresponding to multiple magnetic configurations. Since the meta-stable spin orders as well as the ground state are usually well stabilized from the self-consistent calculations, one can estimate the energy difference and thereby extract $J$ from the mapping onto the spin Hamiltonian. Another way of calculating $J$ from first-principles is often called magnetic force theorem (MFT) \cite{oguchi_magnetism_1983,liechtenstein_local_1987,antropov_spin_1996,katsnelson_first-principles_2000,bruno_exchange_2003,han_electronic_2004,wan_calculation_2006,szilva_interatomic_2013,bessarab_calculations_2014,pi_calculation_2014,steenbock_greens-function_2015,korotin_calculation_2015,kvashnin_microscopic_2016}. In MFT, the interaction strength is calculated as a response to small spin rotation or tilting from the ground state spin density \cite{liechtenstein_local_1987}, which is quite similar with the idea of neutron scattering measurement (Fig. \ref{Figure_1}).

Since its first suggestion by Liechtenstein and co-workers, MFT formalism has been continually developed until quite recently \cite{bruno_exchange_2003,han_electronic_2004,wan_calculation_2006,szilva_interatomic_2013,bessarab_calculations_2014,pi_calculation_2014,steenbock_greens-function_2015,korotin_calculation_2015,kvashnin_microscopic_2016}. While it was originally and largely developed within LMTO (linearized muffin-tin orbital)-based methods, MFT is also extended to LCPAO (linear combination of pseudo-atomic  orbitals) \cite{han_electronic_2004}  and the plane-wave basis method \cite{korotin_calculation_2015}. It is further developed to be applicable to the multipolar exchange \cite{pi_calculation_2014} and the non-collinear spin \cite{szilva_interatomic_2013, bessarab_calculations_2014} interactions. MFT has several advantages over the total energy approach to calculate $J$ while the results of both methods are dependent on the computation details as expected. In the case of the muffin-tin orbital method, the underlying basis-set dependence has been investigated by Kvashnin \textit{et al}. very recently \cite{kvashnin_exchange_2015}.

The purpose of this paper is to establish the reliability of MFT and to pursue its further applicability. First, we examine the $J$ dependence on the numerical parameters within LCPAO formalism (just as Ref. \onlinecite{kvashnin_exchange_2015} does for LMTO). Our comparative and systematic investigation provides the quantitative estimation of possible deviations caused by such details in numerics for both MFT and total energy method. In order to explore its applicability, we examine the MFT results of NiO, bcc Fe, and fcc Ni based on the different spin orders (namely, the ground state and meta-stable spin orders). Our results show that MFT can be used for local moment systems even when the ground state magnetic configuration is unknown. Finally, we extend our MFT to have an orbital resolution. By applying it to several different Fe-based superconductors, we show that the intra- and inter-orbital magnetic interactions in single-stripe Fe arsenides (\ch{LaFeAsO}, \ch{BaFe_{2}As_{2}} and NaFeAs) are clearly distinct from those in the double-stripe chalcogenides (\ch{FeTe}). The inter-atomic AFM (antiferromagnetic) interactions contain some portion of inter-orbital FM (ferromagnetic) couplings and vice versa. Our results demonstrate the usefulness of MFT for studying the complex magnetic materials with multi-band multi-orbital nature.

\section{Computation Details}

We performed density functional theory (DFT) calculations within GGA-PBE (generalized gradient approximation)\cite{perdew_generalized_1996} and GGA$+U$ \cite{anisimov_band_1991,dudarev_electron-energy-loss_1998,han_o_2006}.
Vanderbilt-type norm-conserving pseudopotential\cite{vanderbilt_soft_1990} with a partial-core correction \cite{louie_nonlinear_1982} were used to replace the deep core potentials by norm-conserving soft potentials in a separable form \cite{blochl_generalized_1990}. All of the electronic structure and total energy calculations were carried out with `OpenMX' software package \cite{openmx} which is based on the LCPAO. Throughout the manuscript, we used a notation of `$sN_s~pN_p~dN_d~fN_f$' to represent our basis set choice for a given atom. For example, `$s3p2d1$' denotes that three $s$, two $p$ and one $d$ orbitals were taken as a basis set. If not mentioned otherwise,  R$_{\rm cut}$ (cutoff radii) = 6.0, 5.0, 7.0, 7.0, 9.0, 10.0 and 7.0 a.u. (atomic unit) for transition metals (Mn, Fe, Co, Ni), O, As, Te, Na, Ba, and La, respectively, and multiple zeta and polarization orbitals were used. Further details of our basis set theory can be found  in Ref.~\onlinecite{ozaki_variationally_2003,ozaki_numerical_2004}. For transition metal mono-oxides (MnO, FeO, CoO, and NiO) and elemental transition metals (bcc Fe, fcc Ni and hcp Co), 23$\times$23$\times$23 and 20$\times$20$\times$20 ${\bf k}$ points were used, respectively. For \ch{FeTe}, NaFeAs,  $\ch{BaFe_{2}As_{2}}$ and \ch{LaFeAsO}, we used ${\bf k}$ meshes of 20$\times$20$\times$7 per formula unit. For Fe-based superconductors we used LDA (local density approximation) \cite{ceperley_ground_1980, perdew_self-interaction_1981}. A quantitatively important issue is about the charge analysis scheme. In our calculation, we took Mulliken charge at each transition-metal site.

Our implementation of MFT followed the original formalism by Liechtenstein and co-workers  \cite{liechtenstein_exchange_1984,liechtenstein_local_1987,antropov_spin_1996,antropov_exchange_1997} as extended to non-orthogonal LCPAO method \cite{han_electronic_2004}. In this study, we further refined it to have orbital resolution (Appendix A). Throughout the manuscript we used the following convention for spin Hamiltonian, 
\begin{equation} \label{Eq_Heisenberg_spin_Hamiltonian}
 H=-\sum_{i\neq j} J_{ij} {\bf e}_i\cdot{\bf e}_j.
\end{equation}
where ${\bf e}_{i,j}$ refers to the unit spin vectors of atomic site $i$ and $j$.

For DFT$+U$ calculations, we estimated the material-dependent $U$ parameters from cRPA (constrained random phase approximation) calculations \cite{aryasetiawan_frequency-dependent_2004, sasioglu_effective_2011, sasioglu_strength_2012, sasioglu_textitab_2013}. The current implementation based on `{\tt ecalj}' software package \cite{ecalj-1} was successfully applied to various cuprate superconductors \cite{jang_direct_2016}. We used $U$ $=$ 5.25, 4.75, 4.50 and 4.25 eV for NiO, CoO, FeO and MnO, respectively.

\section{Numerical parameter dependence}\label{sec:parameter_J}

It is important to understand quantitatively the dependence of calculated $J$ on the numerical parameters since there is always some degree of freedom or uncertainty in their choice. However, the detailed examination of basis set dependence has been made only very recently for LMTO-based MFT formalism  \cite{kvashnin_exchange_2015}. In this Section, we analyze the behaviors of $J$  for both MFT and total energy method within LCPAO formalism.

\begin{figure}[!th]
    \includegraphics[width=0.95\linewidth,angle=0]{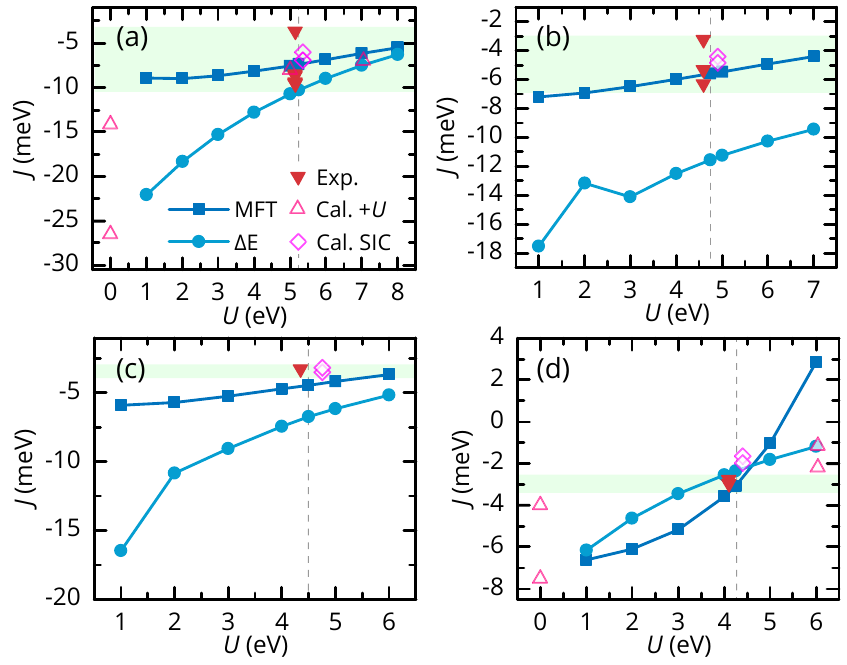}
    \caption{ (Color online) The calculated strongest (second nearest-neighbor) exchange interactions, $J_2$,
    in transition-metal monoxides as a function of $U$; (a) NiO, (b) CoO, (c) FeO, and (d) MnO. Both results from
    MFT and total energies (`$\Delta$E') are presented. The previous experimental results are plotted with filled triangles (red). The previous DFT+$U$ (empty triangles) and SIC (self-interaction correction) DFT results (empty diamond) are also presented. The vertical dotted lines are the cRPA $U$ values.
    The experimental and SIC-DFT results are plotted near the cRPA $U$ values. The shaded region (light green) shows the range of the experimental results.
    For the previous results of experiments and calculations see Ref.~\onlinecite{smart_magnetism_1963,bartel_exchange_1971,bartel_stability_1972,hutchings_measurement_1972,shanker_analysis_1973} and Ref.~\onlinecite{oguchi_band_1983,kodderitzsch_exchange_2002,han_o_2006,kotani_spin_2008,fischer_exchange_2009,kotani_spin_2008,jacobsson_exchange_2013} for NiO; Ref.~\onlinecite{sakurai_crystal_1968,rechtin_long-range_1972,tomiyasu_magnetic_2006} and Ref.~\onlinecite{fischer_exchange_2009} for CoO; Ref.~\onlinecite{kugel_low-energy_1978} and Ref.~\onlinecite{fischer_exchange_2009} for FeO; Ref.~\onlinecite{lines_antiferromagnetism_1965,kohgi_inelastic_1972,pepy_spin_1974} and Ref.~\onlinecite{solovyev_effective_1998,jacobsson_exchange_2013,fischer_exchange_2009} for MnO, respectively.
      }
    \label{Fig_UvsJ_TMO} 
\end{figure}

Let us start with the dependence on a physical parameter, $U$. Figure \ref{Fig_UvsJ_TMO} shows the calculated $J_2$, the second neighbor interaction of transition-metal monoxides, which is known as the major magnetic coupling in this series of materials. The results of MFT and total energy method are compared with previous calculations and experiments. First, it is noted that AFM interaction is gradually reduced as $U$ increases, which is a well-known feature as expected  from the superexchange nature of this interaction \cite{kanamori_crystal_1960,goodenough_magnetism_1963}. Since the determination of material-dependent $U$ is still challenging, one needs to be careful when comparing the calculated $J$ with experiments. It is also noted that, in comparison to experimental values, the best agreement can be reached at different $U$ in MFT and total energy comparison (See Fig. \ref{Fig_UvsJ_TMO}). The vertical dotted lines correspond to our cRPA $U$ values, at which both calculation methods basically give the reliable estimation. Hereafter, we use the cRPA $U$ values to examine the other numerical parameter dependence.

\begin{figure}[!ht]
    \begin{center}
        \includegraphics[width=0.95\linewidth,angle=0]{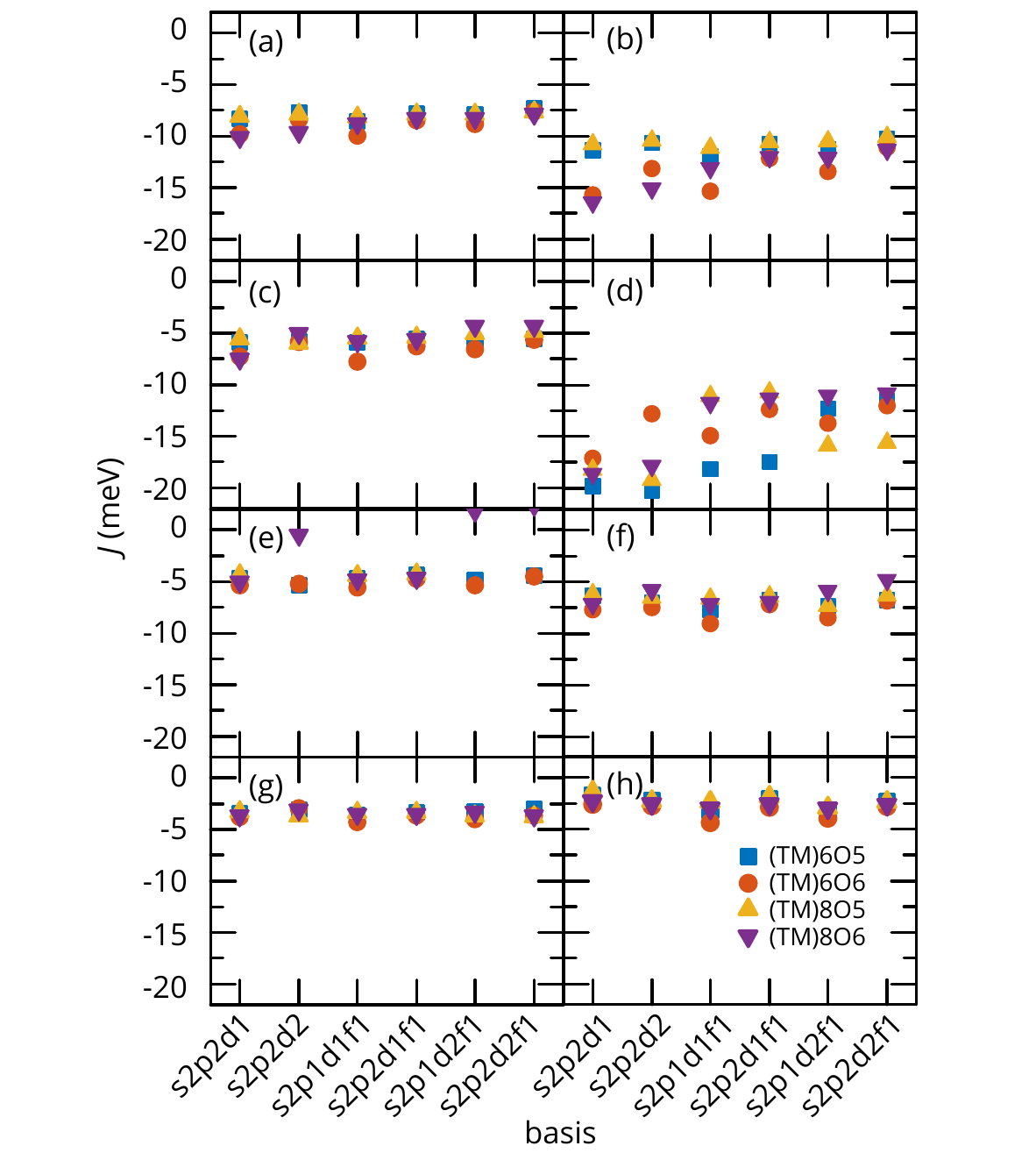}
        \caption{ (Color online) The basis-set dependence of the calculated $J_2$ for 
        	(a, b) NiO, (c, d) CoO, (e, f) FeO,  and (g, h) MnO. 
            The results from MFT (left panels; (a),(c),(e),(g))
            and total energy comparison (right panels; (b),(d),(f),(h)) are presented.
            The $\rm{R_{cut}}$ of transition metals and oxygen are given by the number
            on the right side of `TM' and `O', respectively, in the atomic unit.
            For example, `TM6O5' means that $\rm{R_{cut}}$=6.0 and 5.0 a.u. were used
            for transition metal and oxygen, respectively. 
            \label{Fig_TMO_basis_test}}
    \end{center}
\end{figure}

One characteristic feature of many local-orbital-based methods is that the control of basis set is less straightforward than the plane-wave method. In particular, the number of basis orbitals and the cutoff radius is always somewhat arbitrary while a systematic way of extension is basically available \cite{ozaki_variationally_2003}. Figure~\ref{Fig_TMO_basis_test} presents the results of basis-set dependence in which $J_2$ are calculated for transition-metal monoxides with 24 different basis-set choices. It is clearly seen that, from both MFT and total energies, the $J_2$ are reasonably well estimated in comparison to the reported experimental values. The difference caused by the basis sets is about $\sim$3 meV in the case of MFT while it is slightly larger in the total energy method. The deviation in between different R$_{\rm cut}$ mostly decreases as the number of orbitals increases. Considering the ambiguities in the experiments and other computation details, we conclude that this is in an acceptable range; see the similar size of deviations in both experiments \cite{lines_antiferromagnetism_1965,bartel_exchange_1971,bartel_stability_1972,hutchings_measurement_1972,kohgi_inelastic_1972,shanker_analysis_1973,pepy_spin_1974,rechtin_long-range_1972,sakurai_crystal_1968,smart_magnetism_1963,kugel_low-energy_1978,tomiyasu_magnetic_2006} and calculations \cite{solovyev_effective_1998, jacobsson_exchange_2013,fischer_exchange_2009,kotani_spin_2008,han_o_2006,oguchi_band_1983,kodderitzsch_exchange_2002}, which reflects the various ambiguities in the measurement, fitting, and computation details. The fact that the deviations caused by different basis sets are smaller in MFT than in total energy comparison likely indicates that the ground state electronic structure is robustly well described by any reasonable choice of basis orbitals.

\begin{figure}[!h]
    \includegraphics[width=0.95\linewidth,angle=0]{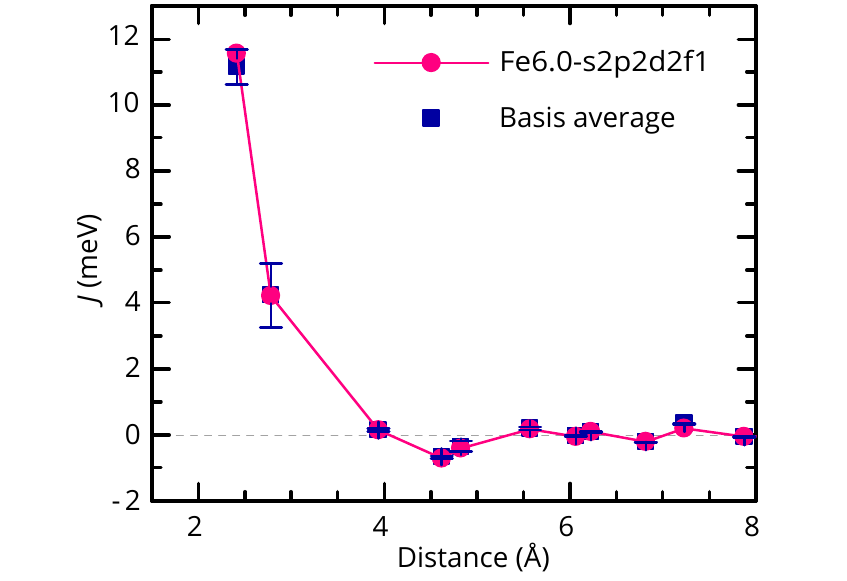}
    \caption{
        The calculated $J$ for bcc Fe as a function of inter-atomic distance. The blue squares represent the average values calculated by using 12 different basis-set choices (same with the choice of Fig.~\ref{Fig_TMO_basis_test}). The blue error bars indicate the standard deviations of using different basis sets. The red circles are from $s2p2d2f1$ and $\rm{R_{cut}}$=6.0 a.u..
        \label{Fig_Metal_basis_test}}
\end{figure}

Figure~\ref{Fig_Metal_basis_test} shows the basis-set dependence result for a metallic system, namely bcc Fe. Considering the long-range feature, we only present MFT in this case. The red circles show the results of R$_{\rm cut}$=6.0 a.u.~ and $s2p2d2f1$ while the blue squares represent the average of 12 different basis-set choices. Importantly, the basis-set dependence is not significant also for metallic case (as far as any reasonable set is chosen). We performed the same examination for fcc Ni and hcp Co (not shown) whose results also support this conclusion.

It is instructive to compare our results with a similar study of using LMTO-based method \cite{kvashnin_exchange_2015}. Recently, Kvashnin et al. examined the LMTO-basis-set dependence of MFT exchange constant for hcp Gd. The use of L\"owdin orthonormalized LMTO (called `ORT') and muffin-tin head projected orbitals (`MTH') \cite{kvashnin_exchange_2015} give the different $J$ values by up to 1.5 meV, which are comparable with what we found in Fig.~\ref{Fig_Metal_basis_test}. Also, this amount of differences are typically observed in the experiments  \cite{bartel_stability_1972,bartel_exchange_1971,hutchings_measurement_1972,kohgi_inelastic_1972,lines_antiferromagnetism_1965,shanker_analysis_1973,pepy_spin_1974,rechtin_long-range_1972,sakurai_crystal_1968,smart_magnetism_1963,kugel_low-energy_1978,tomiyasu_magnetic_2006} as well as calculations performed with different computation details (other than the basis sets; such as exchange-correlation functional \cite{solovyev_effective_1998,kotani_spin_2008,fischer_exchange_2009}).

\section{Magnetic force calculated by non-collinear spin total energies}

\begin{figure}[!h]
    \centering
    \includegraphics[width=0.95\linewidth]{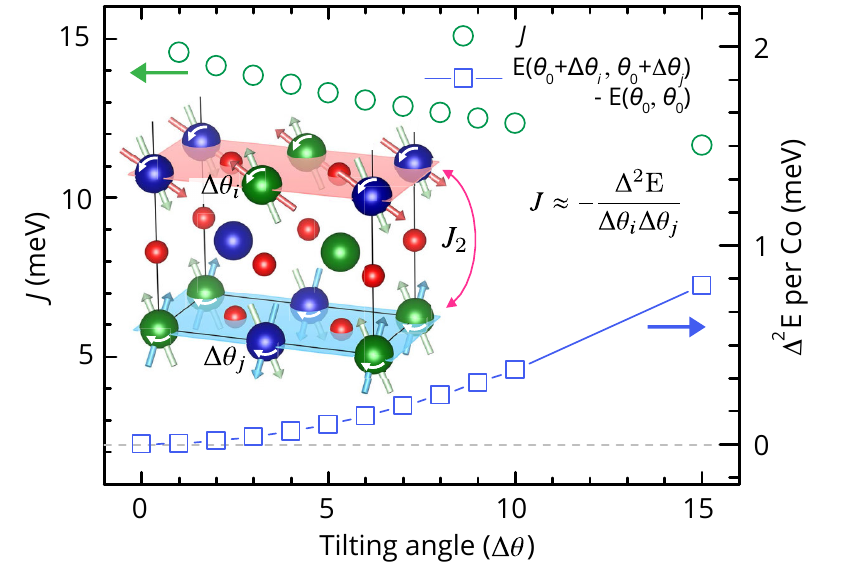}
    \caption{
The calculated total energies (per formula unit) as a function of rotation angle $\theta$ with respect to the easy axis ($\theta_0 = 155\degree$, $\phi_0 = 45\degree$). The blue squares and green circles present the total energies and $J_2$, respectively. The tilting is designed to extract $J_2$ only with no contribution from $J_1$ and others. The unit cell structure and tilting pattern is presented in the inset where the green, blue, and red spheres represent the up-spin, down-spin Co atoms, and oxygen atoms, respectively. The green arrows on the Co sites show the easy axis direction.
        }
    \label{fig:figncrot}
\end{figure}

An interesting way of comparing the magnetic force theorem with the total energy method is to estimate the energy difference caused by small spin angle tilting which is feasible within non-collinear spin calculation. This is an entirely different computation approach including the effect of spin-orbit interaction, but conceptually it is equivalent to MFT.

Figure \ref{fig:figncrot} shows the calculated total energies (blue square) at small angle tiltings with respect to the magnetic easy axis ($\theta_0$=155$\degree$, $\phi_0$=$45 \degree$) for CoO \cite{roth_magnetic_1958,herrmann-ronzaud_equivalent_1978,jauch_crystallographic_2001}. The supercell geometry  is shown in the inset, which is the smallest one to extract $J_2$ with no other contribution. Figure~\ref{fig:figncrot} also presents the $J_2$ values (green circles) from the magnetic forces calculated by the non-collinear spin total energies, which are reasonably well compared with the results of MFT ($J_2$=5.6 meV) and collinear-spin total energy comparison ($J_2$=11.6 meV) presented in  Fig.~\ref{Fig_TMO_basis_test}. 
The difference of $\sim$5 meV reflects the methodological differences as well as spin-orbit coupling effect.

\section{MFT based on meta-stable spin configurations}\label{sec:MFT_metastable}

\begin{figure}[!ht]
    \includegraphics[width=0.95\linewidth]{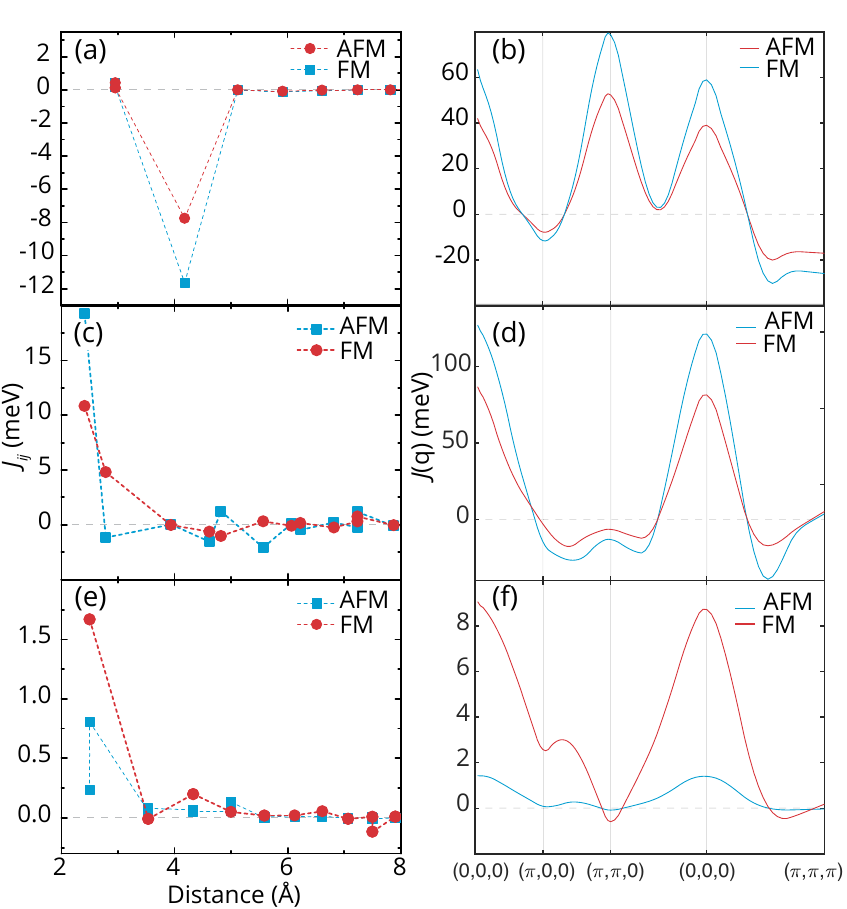} %
    \caption{ The calculated $J$ for (a, b) NiO, (c, d) bcc Fe, and (e, f) fcc Ni.
        The real space $J_{ij}(r)$ and the momentum space $J_{ij}{\rm{(q)}}$ are presented in
        (a), (c), (e) and (b), (d), (f), respectively. $J_{ij}(r)$ is calculated as a function of inter-atomic distance. For $J_{ij}{\rm{(q)}}$, $i, j$ are set to represent the strongest neighbor interactions. The calculations are based on both FM and AFM spin order. We used the red colored lines to represent the ground state spin configurations, while the blue lines refer to the results from metastable magnetic configurations.
        \label{Fig_Meta_TMO} }
\end{figure}

One notable feature of MFT is that it only makes use of the electronic structure. In other words, the calculation of $J$ based on MFT does not require total energy information. This can be a great advantage for a practical purpose since calculating total energy is not straightforward in many electronic structure calculation methods. For example, GW method often provides the improved description of electronic structure, but it is not easy to calculate total energy in both self-consistent and non-self-consistent version of GW (for the related discussion, see also Ref.~\onlinecite{kutepov_ground-state_2009,caruso_unified_2012}). In another advanced scheme, namely DFT+DMFT (dynamical mean field theory), total energy calculation is not straightforward although there are some recent successful reports \cite{leonov_computation_2010,park_total_2014,haule_free_2015}. We also note that GW and DFT+DMFT are computationally demanding, and therefore even if the total energy calculation is formally feasible, the consideration of larger unit cells to describe the multiple magnetic configurations is often impractical. In this regard, MFT can be a better choice. Further, the magnetic ground state is sometimes unknown due to the practical reasons such as the sample size and quality issues.

Considering these features, in this Section, we pursue the further applicability of MFT. Figure~\ref{Fig_Meta_TMO} presents the calculation results of three different materials ranging from the strongly correlated insulating NiO (a, b) to the weakly or moderately correlated metals, Fe (c, d) and Ni (e, f). In particular, we performed MFT calculation based on the meta-stable phase as well as the ground state spin order. Figure~\ref{Fig_Meta_TMO}(a) and (b) show that the two results estimated from the ground state AFM and meta-stable FM configuration of NiO are not much different from each other. The real space coupling $J_{ij}(r)$ (Fig.~\ref{Fig_Meta_TMO}(a)) clearly shows that the second neighbor interaction ($J_2$) is dominant while the long-range interactions are negligible as expected. Interestingly, $J$'s from AFM and FM are quite similar also in the momentum space ($J_{ij}(q)$; Fig.~\ref{Fig_Meta_TMO}(b)). The same features are found for MnO, FeO, and CoO (not shown). It means that the magnetic interaction can be reasonably estimated based on {\it any} stable magnetic solution. The possible error range is a few meV, which is probably acceptable considering other possible ambiguities as discussed in Sec. \ref{sec:parameter_J}. This result reflects that, for strongly localized moment system, the magnetic response induced by small perturbations is not much dependent on the long-range spin order type. This feature can be useful especially when the total energy calculation is not feasible and/or the ground state configuration is unknown.

In case of less localized moments, this feature is gradually weakened. In Fig. \ref{Fig_Meta_TMO}(c) and (d), the results of bcc Fe are presented where the similarity between the results from AFM (meta-stable phase) and FM (the real ground state) is less clear. However, in both $J(\rm{r})$ and $J(\rm{q})$, the two results are still quite consistent. On the other hand, in the least correlated case of Ni (Fig. \ref{Fig_Meta_TMO}(e) and (f))\cite{lichtenstein_finite-temperature_2001}, two results from FM and AFM become significantly different. This result implies that for weakly correlated systems far from the local moment picture, MFT should be based on the real ground state solution, which limits its applicability.

\section{Orbital-resolved magnetic interactions}\label{sec:orbital_resolution} 

Another advantage of using MFT is to have additional information for magnetic interactions. Here we note that while the typical $J$ values are given in numbers either estimated from calculated total energies or experiments, the linear response procedure within MFT can provide the orbital-dependent response to the spin angle tilting \cite{korotin_calculation_2015,kvashnin_microscopic_2016}. This kind of approach can provide further insights especially for the complex magnetism of multi-band nature.
In this section, we apply the newly-implemented technique of orbital-resolved $J$ to Fe-based superconductors. While our work is based on LCPAO formalism (the details are given in Appendix A), it is straightforward to extend it to other basis methods, {\it e.g.}, using Wannier functions \cite{korotin_calculation_2015,kvashnin_microscopic_2016}.

The magnetic interactions in Fe-based materials are certainly a key to understand the superconductivity. Due to their multi-band multi-orbital nature, however, it is highly non-trivial to analyze those magnetic interactions and to make comparisons for the characteristic material dependency. Recently, both spin and orbital fluctuation have been intensively discussed  \cite{kuroki_unconventional_2008,yildirim_origin_2008,kuroki_unconventional_2008,bang_possible_2008,kuroki_pnictogen_2009,kontani_orbital-fluctuation-mediated_2010,onari_self-consistent_2012,fernandes_what_2014,fanfarillo_spin-orbital_2015,hosono_iron-based_2015,dai_spin-fluctuation-induced_2015,yamakawa_nematicity_2016}.

The calculation result of orbital-dependent magnetic interactions for \ch{LaFeAsO} is
\[
J_{1a}= \quad \begin{blockarray}{rrrrrr}
d_{z^2}  & d_{x^{2}y^{2}} & d_{xy} & d_{xz} & d_{yz} &  \\
\begin{block}{ (rrrrr){l}}
0.10  & -1.59 & 0.00  & -0.05 & 0.00   & \quad d_{z^2} \\
-1.59 & 1.65  & 0.00  & -2.48 & 0.00   & \quad d_{x^{2}y^{2}} \\
0.00  & 0.00  & -0.76 & 0.00  & -1.83  & \quad d_{xy}   \\
-0.05 & -2.48 & 0.00  & -1.20 & 0.00   & \quad d_{xz}   \\
0.00  & 0.00  & -1.83 & 0.00  & 1.13   & \quad d_{yz}  \\
\end{block}
\end{blockarray}
\]

\[
J_{1b}= \quad \begin{blockarray}{rrrrr}
\begin{block}{ (rrrrr)}
-0.01 &	-1.32	&  0.00	 & 0.00	 & -0.45 \\
-1.32 & 2.13	&  0.00	 & 0.00	 & -2.55 \\
0.00  & 0.00	&  2.15	 & -1.25 & 	0.00 \\
0.00  & 0.00	&  -1.25 & 4.40	 &  0.00 \\
-0.45 & -2.55	&  0.00	 & 0.00	 &  0.60 \\
\end{block}
\end{blockarray}
\]
where $J_{1a}$ and $J_{1b}$ refers to the nearest-neighbor interactions along $a$ (AF ordering) and $b$ (FM ordering) axis, respectively (for our notation see, for example, Ref.~\onlinecite{han_anisotropy_2009,han_doping_2009}). It is noted that each constant $J$ is now expressed as a 5$\times$5 matrix whose matrix elements represent the interactions in between two given orbitals. The conventional value of $J$ corresponds to the sum of all these matrix elements. In fact, the summations give the good agreement with previous results \cite{han_anisotropy_2009}. Note that, while the interaction $J_{1a}$ is AFM, some of its orbital components are FM. 
It means that through some orbital channels the moments interact ferromagnetically. This unique information provided by orbital-resolved MFT can be useful for the analysis of complex multi-orbital systems.

We found that the same interaction profiles for \ch{BaFe_{2}As_{2}} and \ch{NaFeAs}. Namely, their $J_{1a,1b}$ matrices are given in the identical form with those of \ch{LaFeAsO}; the same signs for the major components and the zero values at the same matrix elements. The off-diagonal inter-orbital interactions are the major AF couplings. The only difference is the relative strengths of each orbital-dependent coupling (matrix element). This finding strongly indicates that magnetic interactions in these `single-stripe' materials are in fact quite similar even at the orbitally decomposed level.

On the other hand, the interactions of `double-stripe' \ch{FeTe} are notably different:
\[
J_{1a}= \quad \begin{blockarray}{rrrrr}
\begin{block}{ (rrrrr)}
0.06  & -0.97	 & -0.10 & 	-0.25 &  -0.03  \\ 
-0.97 & 0.05	 & 0.00	 & -1.09  &  -0.02  \\
-0.10 & 0.00	 & 3.64	 & -0.26  &  -2.26  \\
-0.25 & -1.09	 & -0.26 & 	0.26  &  0.00   \\
-0.03 & -0.02	 & -2.26 & 	0.00  &  2.37   \\ 
\end{block}
\end{blockarray}
\]

\[
J_{1b}= \quad \begin{blockarray}{rrrrr}
\begin{block}{ (rrrrr)}
0.00    &  -0.86  & -0.10	 & 0.01	 & -0.19 \\
-0.86	&  0.11	  & 0.08	 & 0.03	 & -1.03 \\
-0.10	&  0.08	  & 2.38	 & -2.48 & 	0.00 \\
0.01	&  0.03	  & -2.48	 & 2.77	 & -0.02 \\
-0.19   &  -1.03  & 0.00	 & -0.02 & 	0.22 \\
\end{block}
\end{blockarray}
\]
where most of $J_{1a}$ and $J_{1b}$ matrix elements are non-zero. Our result reflects the longer range nature of Te-$p$ orbitals and clearly shows that FeTe forms a different class distinctive from LaFeAsO and other single-stripe materials \cite{subedi_density_2008,han_anisotropy_2009,miyake_comparison_2010}.

We emphasize that the orbital-resolved magnetic interaction by MFT has a great potential to study the complex magnetism by providing the unique information that is not accessible from other computation techniques and experiments.

\section{Summary}
We investigated magnetic force linear response theory from the point of view of accuracy, reliability, and applicability. Within LCPAO formalism, the error range caused by the choice of basis set is carefully examined, which shows that the numerical parameter dependence is small enough to be compared with experiments and other computation schemes. Magnetic force calculation of $J$ based on the non-collinear total energy results is investigated. To explore its applicability, we examined the $J$ values calculated from different magnetic solutions. For strongly localized moments, the results do not much depend on the magnetic order, and therefore the magnetic interaction can be estimated basically from any stable spin density profile. For less localized systems such as metallic magnets, the dependence becomes noticeable; however, the magnetic force calculation can still give useful information. Finally, we extend the formalism to have an orbital resolution from which the coupling constant $J$ is expressed not by a single number but by a matrix. By applying to Fe-based superconductors, we demonstrate its potential capability to study complex multi-orbital magnetism.

\section{Acknowledgments}
We thank Prof. Kuroki for helpful discussion.
This work was supported by Basic Science Research Program through the National Research Foundation of Korea (NRF)
funded by the Ministry of Education(2017R1D1A1B03032082).

\appendix*
\section{Calculating orbital-resolved exchange parameters}

To calculate the orbital-decomposed $J$ within non-orthogonal LCPAO basis method,
we extend our formalism of Ref.~\onlinecite{han_electronic_2004}.
Namely, the exchange parameters are calculated based on MFT
derived from multiple scattering theory \cite{lloyd_multiple_1972,
oguchi_band_1983,oguchi_magnetism_1983,liechtenstein_exchange_1984,liechtenstein_local_1987,udvardi_first-principles_2003,han_electronic_2004,mazurenko_weak_2005,ebert_anisotropic_2009,antropov_exchange_1997}:
\begin{equation} \label{Eq_Jij_momentumspace}
J_{ij}({\bf{q}} ) = \frac{1}{\pi} {\rm{Im}} \int_{}^{}  
\int_{}^{\epsilon_{\rm{F}}}  d{\rm{k}} \, d\epsilon  
\rm{\, Tr}[
V_{{\bf{k}},i}^{\downarrow \uparrow } {\mathbf{G}}_{{\bf k},ij}^{\uparrow\uparrow}   
V_{{\bf{k+q}},j}^{\uparrow \downarrow}  {\mathbf{G}}_{{\bf k+q},ji}^{\downarrow\downarrow} 
].
\end{equation}
Here the orbital index $l$ can be included:
\begin{equation} \label{Eq_green_DFT}
    \mathbf{G}^{\uparrow \uparrow}_{l_1 l_2 ,{\bf{k}},ij} = \sum_{n}^{} \frac{ \ket{\psi_{l_1 ,{\bf{k}},i}^{\uparrow}} \bra{\psi^{\uparrow}_{l_2  ,{\bf{k}},j}} }{z-\epsilon_{\uparrow n,{\bf{k}}}   + i\eta}.
\end{equation}
Among four orbital indices ($l_1$, $l_2$, $l_3$, $l_4$) in Eq.~(\ref{Eq_Jij_momentumspace}),
$l_1$ and $l_3$ belong to a site $i$ while $l_2$ and $l_4$ to $j$.
Thus, $J_{ ij  }^{l_1,l_2}$ is given by
\begin{equation}\label{Eq_orbital_J}
J_{ ij  }^{l_1,l_2} ( {\bf{q} } ) =\sum_{l_3,l_4}^{} J_{ij}^{l_1, l_2, l_3, l_4} ( {\bf{q} } ) .
\end{equation}
The estimation is straightforward from our LCPAO Hamiltonian ${\bf{H}}_{l_3 l_1,{\bf k},i}^{\uparrow \uparrow }$, ${\bf{H}}_{l_3 l_1,{\bf k},i}^{\downarrow \downarrow }$, and within the collinear spin formalism, the off-diagonal parts (${\bf{H}^{\downarrow\uparrow}}$ and ${\bf{H}^{\uparrow\downarrow}}$) are all zero.
The on-site exchange energy is
\begin{equation}\label{Eq_Vdef}
V_{l_3 l_1,{\bf{k}},i}^{\downarrow  \uparrow } = {\frac{1}{2}}( {\bf{H}}_{l_{3} l_{1},{\bf k},i}^{\downarrow  \downarrow } - {\bf{H}}_{l_{3} l_{1},{\bf k},i}^{\uparrow \uparrow } ).
\end{equation}

Reorganizing the trace operations for Eq.~(\ref{Eq_Jij_momentumspace}) and (\ref{Eq_green_DFT}), the final expression is given by 
\begin{equation}\label{Eq_J_orbital_momentum}
    \begin{aligned}
J_{ ij }^{l_1,l_2} ({\bf q}) = \sum_{l_3,l_4}^{}\sum_{n,m}^{} \sum_{{\bf k}}^{}{f(\epsilon_{\uparrow n, {\bf k} })-f(\epsilon_{\downarrow m,{\bf k+q} })
\over{\epsilon_{\uparrow m,{\bf k+q} }-\epsilon_{\downarrow n,{\bf k} } +i\eta}}  \times\\
{
\braket{\psi_{l_3 ,{\bf{k+q}},i }^{\downarrow}  |V_{l_3 l_1, {\bf{k}},i}^{\downarrow \uparrow }|  \psi_{l_1, {\bf{k}},i}^{\uparrow }}
}
\times \\
\nonumber
\braket{\psi_{l_2, {\bf{k}},j}^{\uparrow}   |V_{l_2 l_4, {\bf{k+q}},j}^{\uparrow \downarrow }|  \psi_{l_4,{\bf{k+q}},j}^{\downarrow }  },
    \end{aligned}
\end{equation}
and the real space value is obtained by Fourier transformation.

%


\end{document}